\documentclass[letter,twocolumn]{jpsj3}
\usepackage{txfonts}
\usepackage{graphicx,bm,color,braket}

\title{Hall Effect in the Vortex Lattice of $d$-Wave Superconductors with Anisotropic Fermi Surfaces}

\author{Wataru Kohno, Hikaru Ueki, and Takafumi Kita}
\inst{Department of Physics, Hokkaido University, Sapporo 060-0810, Japan} %\\
\abst{
On the bas{{i}}s of the augmented quasiclassical theory of superconductivity with the Lorentz force, we study {{the}} magnetic field dependence of the charge distribution due to the Lorentz force in a $d$-wave vortex lattice with anisotropic Fermi surfaces.
{{Owing}} to the competition between the energy-gap and {{Fermi surface}} anisotropies, the charge profile in the vortex lattice changes dramatically with increasing magnetic field because of the overlaps of each nearest vortex-core charge.
In addition, the accumulated charge in the core region may reverse its sign as a function of magnetic field.
This strong field dependence of the vortex-core charge cannot be observed in the model with {{a}}{n} isotropic Fermi surface.

}
\begin{document}
\maketitle
{
It was shown previously\cite{WK} on {{the basis of}} a microscopic theory\cite{Kita01,Ueki} that the vortex-core charge accumulated by the Lorentz force on supercurrents exhibits a characteristic magnetic-field dependence in two-dimensional $s$-wave superconductors.
Specifically, the charge density at the core center $\rho_0$ approximately obeys $\rho_0(H)\propto H(H_{{\rm{c}}2}-H)$ {{owing}} to the competition between the increasing magnetic field $H$ and decreasing pair potential ($H_{{\rm{c}}2}$: upper critical field).
We here investigate how this peak structure and charge profile may be affected by two additional factors, i.e., the {{Fermi surface}} and energy-gap anisotropies.

The charge redistribution due to the magnetic Lorentz force can be described by the Hall coefficient.
It is well known that the {{Fermi surface}} curvature plays a crucial role {{in}} the signs and magnitudes of the normal Hall coefficient.\cite{Ziman}
Another crucial factor {{appears}} for the Hall coefficient of equilibrium superconductors, i.e., the energy-gap anisotropy,
which generally {{introduces}} temperature dependences {{of}} the excitation curvature formed by quasiparticles.
Indeed, it was shown previously by model calculations on a $d_{x^2-y^2}$ pairing that both the Hall coefficient in the Meissner state\cite{Kita09} 
and {{the}} accumulated charge in an isolated vortex core\cite{Ueki} strongly depend on the temperature and may change their signs as a function of temperature.
With these results, one may naturally expect that the two factors are also important {{in the}} magnetic{{-}}field dependences of
charge profiles in vortex-lattice states.
Hence, we here study the charge distribution in the vortex lattice of a clean $d$-wave superconductor with anisotropic Fermi surfaces.

Our calculations are based on the augmented quasiclassical equations with the Lorentz force in the Matsubara formalism\cite{Ueki}.
We can divide them into the standard Eilenberger equations\cite{Kopnin,Parks69,KitaText,Rainer83,LO86} and an electric-field equation\cite{Kita09,Ueki,WK} through an expansion in terms of the dimensionless quasiclassical parameter $\delta\equiv(k_{\rm{F}}\xi_0)^{-1}\ll 1$, where $k_{\rm{F}}$ is the Fermi wave number and $\xi_0$ is the coherence length at $T=0$. These equations can be written as follows{{.}}
\begin{subequations}
\label{Ei}
\begin{align}
&\varepsilon_n f+\frac{1}{2}\hbar{\bm v}_{\rm{F}}\cdot \left(\bm{\nabla}-i\frac{2e{\bm A}}{\hbar}\right)f=\Delta\phi g,\label{Ei1}\\
&\Delta=g_0 \pi  k_{\rm{B}} T \sum_{n=-\infty}^{\infty}\langle f \phi^{*}\rangle_{\rm{F}}\label{Ei3},\\
&\bm{\nabla}\times\bm{\nabla}\times{\bm A}=-i2\pi e\mu_0N(0)k_{\rm{B}}T\sum_{n=-\infty}^{\infty}\langle {\bm v}_{\rm{F}}g \rangle_{\rm{F}}\label{Ei4},\\
&(-\lambda^2_{\rm{TF}}\bm{\nabla}^2+1){{\bm E}}=-i\pi k_{\rm{B}}T{\bm B}\times\sum_{n=-\infty}^{\infty}\left\langle\frac{\partial g}{\partial{\bm p}_{\rm{F}}}  \right\rangle_{\rm{F}},\label{el}
\end{align}
\end{subequations}
with the normalization condition $g={\rm{sgn}}(\varepsilon_n)\sqrt{1-f\bar{f}}$. 
Here $f=f(\varepsilon_n , {\bm p}_{\rm F}, {\bm r})$ and $\bar{f}\equiv f^{*}(\varepsilon_n , -{\bm p}_{\rm F}, {\bm r})$ are anomalous quasiclassical Green's functions, $g_0\ll 1$ is a dimensionless coupling constant responsible for Cooper pairing, $\mu_0$ and $\epsilon_0$ are the vacuum permeability and permittivity, and $\lambda_{\rm{TF}}\equiv\sqrt{\epsilon_0/2e^2N(0)}$ is the Thomas--Fermi screening length, respectively. 
$\langle \cdots \rangle_{\rm{F}}$ denotes {{the}} Fermi surface average normalized as $\langle 1 \rangle_{\rm{F}}=1$ and $\phi=\phi(\bm{p}_{\rm{F}})$ denotes {{the}} gap anisotropy normalized as $\langle |\phi|^2 \rangle_{\rm{F}}=1$. 
Equations (\ref{Ei3}) and (\ref{Ei4}) are the self-consistency equations for the pair potential $\Delta(\bm{r})$ and the vector potential $\bm{A}(\bm{r})${,} {{respectively}}. 
The {{electric}} field $\bm{E}(\bm{r})$ due to the Lorentz force can be obtained from Eq. (\ref{el}), which consists of {{the}} quasiclassical Green's function $g$ and $\bm{B}={\bm\nabla}\times {\bm A}$, namely, the solutions of the Elienberger equations {{(\ref{Ei1}), (\ref{Ei3}) and (\ref{Ei4})}}.
Substituting $\bm{E}$ into Gauss' law $\rho(\bm{r})=\varepsilon_{0}\bm{\nabla}\cdot \bm{E}$, we can find the charge distribution.

For the single-particle energy, we adopt the following dimensionless dispersion for a two-dimensional square lattice used for high-$T_{\rm{c}}$ superconductors \cite{Kontani,Kita09}{{:}}
\begin{align}
\varepsilon_{\bm{p}}=&-2(\cos p_{x}+\cos p_{y})+4t_{1}(\cos p_{x}\cos p_{y}-1)\notag\\
&+2t_{2}(\cos 2p_{x}+\cos 2p_{y}),\label{tight}
\end{align}
with $t_{1}=1/6$ and $t_{2}=1/5$, which forms {{a}} band over $-4\leq\varepsilon_{\bm{p}} \leq4$. 
The structure of the Fermi surface $\varepsilon_{\bm{p}}=\varepsilon_{\rm{F}}$ is determined by the average electron filling per site $n$ $(0\leq n\leq2)$;
Fig.\ \ref{fig1} shows the Fermi surfaces of $n=0.9$ and $n=1.95$ for the single-particle energy of Eq.\ (\ref{tight}).
{{
Each of them is given in the extended zone scheme by a singly connected contour around $(p_{x},p_{y})=(\pi,\pi)$}}.
On the other hand, we express a $d_{x^2-y^2}$-wave symmetric energy gap as $\phi(\bm{p}_{\rm{F}})\propto[(p_{{\rm{F}}x}-\pi)^2-(p_{{\rm{F}}y}-\pi)^2]$ {{$(0\leq p_{x},p_{y}\leq\pi)$}} {{which is}} appropriate for $n\gtrsim0.8$. 

Vortex-lattice solutions of {the} $d$-wave superconductor are constructed by using the Eilenberger equations.
The corresponding vector potential is expressible in terms of the average flux density $\bar{\bm B}=(0,0,\bar{B})$ as ${\bm A}({\bm r})=(\bar{\bm B}\times{\bm r})/2+\tilde{\bm A}({\bm r})$,\cite{Ichioka1,KitaGL} where $\tilde{\bm A}$ describes {{the}} spatial variation of the flux density.
Previous studies {{on}} the vortex lattice configuration\cite{Ichioka2,VFex1,VFex2} show{{ed}} that the square lattice state is more stable than the triangular lattice state over a wide range of fields for the $d_{x^2-y^2}$-wave pairing. 
{{Thus}}, we use the square lattice throughout this analysis.
Functions $\tilde{\bm A}$ and $\Delta$ for the square lattice obey the following periodic boundary conditions:\cite{Klein,Ichioka1,KitaGL}
\begin{subequations}
\label{sym}
\begin{align}
&\tilde{\bm A}({\bm r}+{\bm R})=\tilde{\bm A}({\bm r})\label{sym1},\\
&\Delta({\bm r}+{\bm R})=\Delta({\bm r})\exp\left[i\frac{|e|}{\hbar}{\bar{\bm B}}\cdot\left({\bm r}\times{\bm R}\right)+i\pi n_1n_2\right]\label{sym2},
\end{align}
\end{subequations}
where ${\bm R}=n_1{\bm a}_1+n_2{\bm a}_2$ with $n_1$ and $n_2$ denoting integers, and ${\bm a}_1=a_2(1/2,1/2,0)$ and ${\bm a}_2=a_2(0,1,0)$ are 
the basic vectors of the square lattice with the length $a_2$ determined by the flux-quantization condition {{$({\bm a}_1\times{\bm a}_2)\cdot\bar{\bm B}=h/2|e|$}}.

We follow the numerical procedures shown in {{Ref. \citen{WK}}} to solve the Eilenberger equations with the boundary conditions in {{Eq. (\ref{sym})}}. 
The convergence of the iteration can be checked by monitoring the free energy of the unit cell. 
We confirmed that the free energy decreases as the iteration proceeds, which was stopped when the relative difference between the old and new free energies decreased to below $10^{-5}$.

 \begin{figure}[t]
        \begin{center}
        \vspace{1.3mm}
                \includegraphics[width=0.83\linewidth]{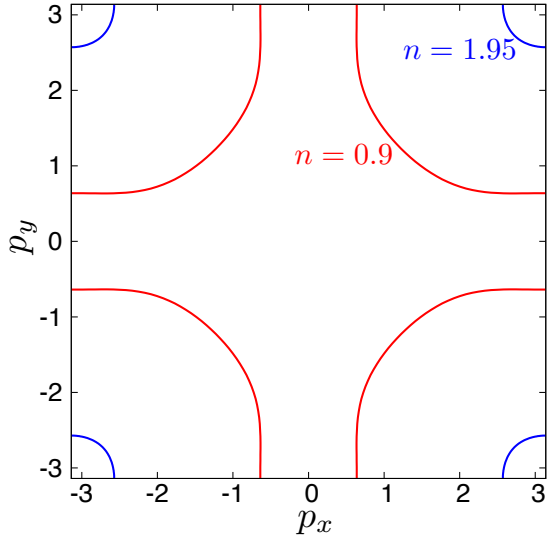}
                \end{center}
                \caption{(Color online) Fermi surfaces of $n=0.9$ (red line) and $n=1.95$ (blue line) for the single-particle energy of Eq.\ (\ref{tight}).}
                \label{fig1}
     \end{figure}
We need to obtain $H_{{\rm{c}}2}$ to investigate the vortex lattice state in the range $H_{{\rm{c}}1}\leq {\mu_0\bar{B}} \leq H_{{\rm{c}}2}$.
In order to {{obtain}} $H_{{\rm{c}}2}$ corresponding to the present model, we derive an {{equation for $H_{{\rm{c}}2}$}}\cite{KA04} {{that}} incorporates the effects of the {{Fermi surface}} and energy-gap anisotropies.
To this end, we transform {{Eqs}}. (\ref{Ei1}) and (\ref{Ei3}) into an algebraic equation by expanding $\Delta$ and $f$ in the basis function using the solutions of the linearized {{Ginzburg--Landau}} equations $\psi_{N,{\bm{q}}}$;
\begin{align}
 \Delta({\bm{r}})&=\sqrt{V}\sum_{N=0}^{N_{{\rm{cut}}}}\Delta_{N}\psi_{N,{\bm{q}}}({\bm{r}}),\label{Lan1}\\
 f({\bm{r}},{\bm{p}}_{{\rm{F}}},\varepsilon_n)&=\sqrt{V}\sum_{N=0}^{N_{{\rm{cut}}}}f_{N}({\bm{p}}_{{\rm{F}}},\varepsilon_n)\psi_{N,{\bm{q}}}({\bm{r}}),\label{Lan2}
 \end{align}
 where $N=0,1,2,\cdots$ denotes the Landau level, $\bm {q}$ is an arbitrary chosen magnetic Bloch vector characterizing the broken translational symmetry of the vortex lattice and specifying the core locations, and $V$ is the volume of the system. 
We subsequently take the normal-state limit $g\to{\rm{sgn}}(\varepsilon_{n})$ and $\bm{A}\to \bar{B}x\bm{\hat{y}}$ in Eq. (\ref{Ei1}), substitute Eqs. (\ref{Lan1}) and (\ref{Lan2}) into Eqs. (\ref{Ei1}) and (\ref{Ei3}), multiply them by $\psi^{*}_{N',{\bm{q}}}$, and perform integrations over ${\bm{r}}$.
Equations ({\ref{Ei1}}) and ({\ref{Ei3}}) are thereby transformed into
 \begin{figure}[t]
        \begin{center}
                \includegraphics[width=0.94\linewidth]{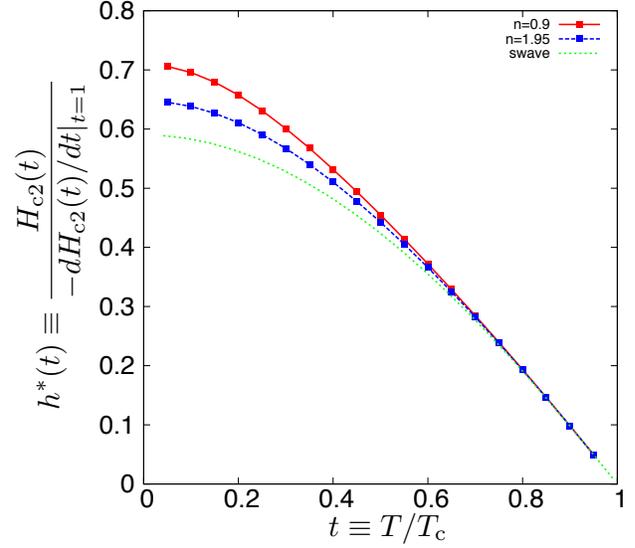}
                \end{center}
                \caption{(Color online) Curves of the reduced critical field $h^{*}(t)$ for the model of a $d$-wave superconductor with n=0.9 (red solid line with points) {{and}} 1.95 (blue dashed line with points) and an $s$-wave superconductor with {{a}} circular Fermi surface (green dotted line).}
                \label{fig2}
     \end{figure}
   \begin{align}
& \sum_{N'}{\mathcal{M}}_{N,N'}f_{N'}=\phi\Delta_{N},\label{ex1}\\
 &\Delta_{N}= g_0\pi k_{\rm{B}} T \sum_{n=-\infty}^{\infty}\langle  f_{N}\phi^* \rangle_{\rm{F}},\label{ex2}
\end{align} 
where $\mathcal{M}$ is a tridiagonal matrix defined by
 \begin{equation}
 {\mathcal{M}}_{N,N'}\equiv|\varepsilon_n|\delta_{N,N'}+\beta^{*}\sqrt{N+1}\delta_{N,N'-1}-\beta\sqrt{N}\delta_{N,N'+1},
\end{equation} 
with $\beta\equiv{\rm{sgn}}(\varepsilon_n)\hbar v_{{\rm{F}}}e^{-i\varphi_{{\bm{v}}_{{\rm{F}}}}}/2\sqrt{2}l_{{\rm{c}}}$, $\varphi_{{\bm{v}}_{{\rm{F}}}}\equiv {\rm{tan}}^{-1}(v_{{\rm{F}}y}/v_{{\rm{F}}x})${{,}} and $l_{\rm{c}}\equiv\sqrt{\Phi_0/2\pi\bar{B}}$.
We now introduce the {{Hermitian}} matrix $\mathcal{K}_{N,N'}\equiv(\mathcal{M}^{-1})_{N,N'}$ and substitute {{Eq.}} (\ref{ex1}) into {{Eq.}} (\ref{ex2}). Then, we obtain the following relation:
 \begin{equation}
\left(\Gamma-\pi g_{0}k_{\rm{B}}T\sum_{n=-\infty}^{\infty}\langle|\phi|^2\mathcal{K}\rangle_{\rm{F}}\right){\bm{\Delta}}\equiv\mathcal{A}{\bm{\Delta}}={\bm{0}},\label{dete}
\end{equation} 
where we define $\Gamma$ as {{an}} ${N_{\rm{cut}}}\times {N_{\rm{cut}}}$ unit matrix and  ${\bm{\Delta}}\equiv(\Delta_0,\Delta_1,\cdots,\Delta_{N_{\rm{cut}}})^{T}$. 
We determined $H_{\rm{{c}}2}$ as the largest solution of the {{equation}}

 \begin{equation}
{\rm{det}}\mathcal{A}(H=H_{{\rm{c}}2})=0.\label{BC2}
\end{equation} 
\begin{figure}[t]
        \begin{center}
                \includegraphics[width=0.9\linewidth]{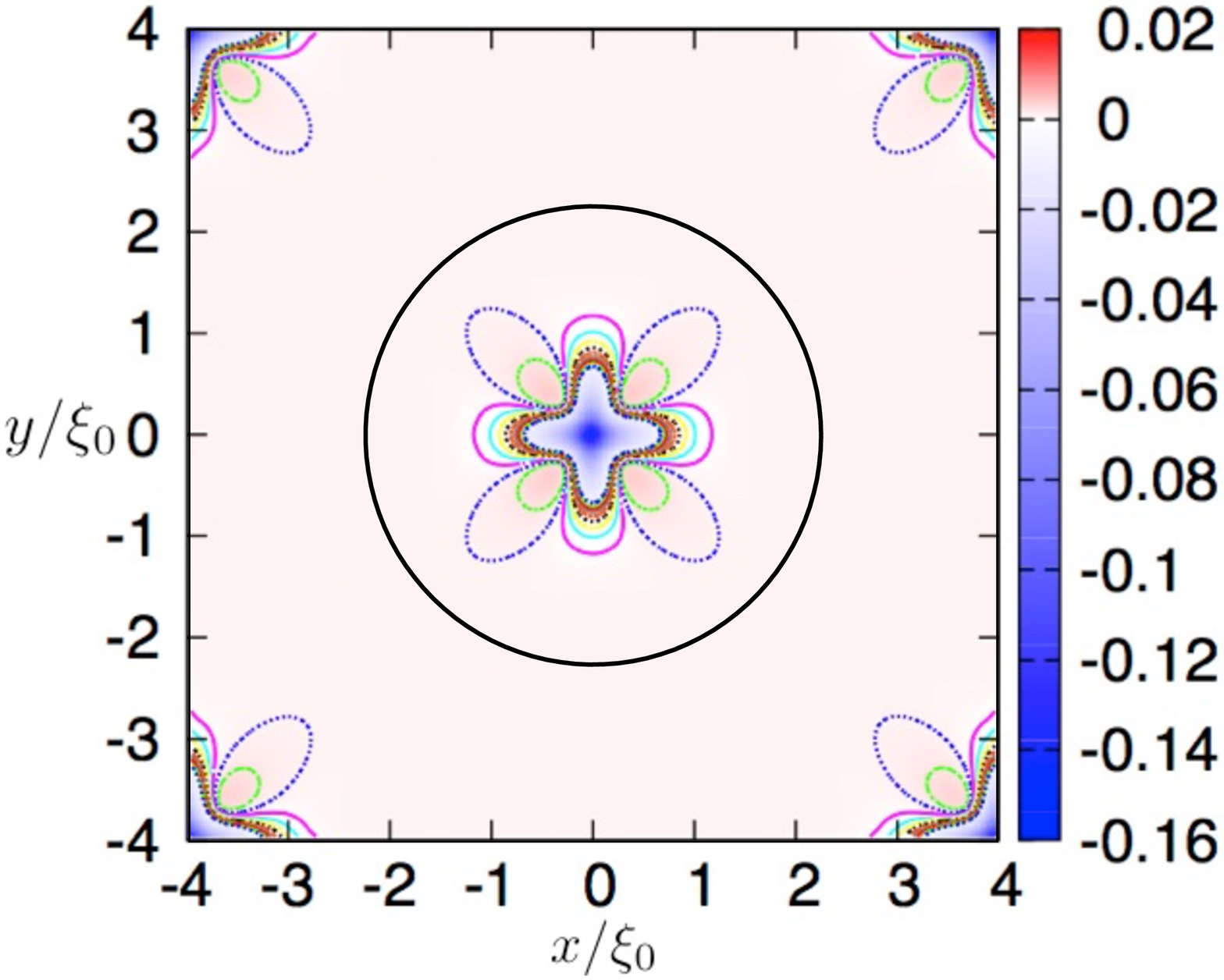}
                \end{center}
\caption{(Color online) Charge density $\rho({\bm r})/\rho_0$ in the vortex lattice calculated for the case of $n=1.95$ and ${{\mu_0\bar{B}/H_{{\rm{c}}2}}}=0.022$ over $-a_2/2\leq x,y \leq a_2/2$, where $a_2/\xi_0=8.0$. }
                \label{fig3}
                \end{figure} 
                \begin{figure}[t]
                \begin{center}
                \includegraphics[width=0.9\linewidth]{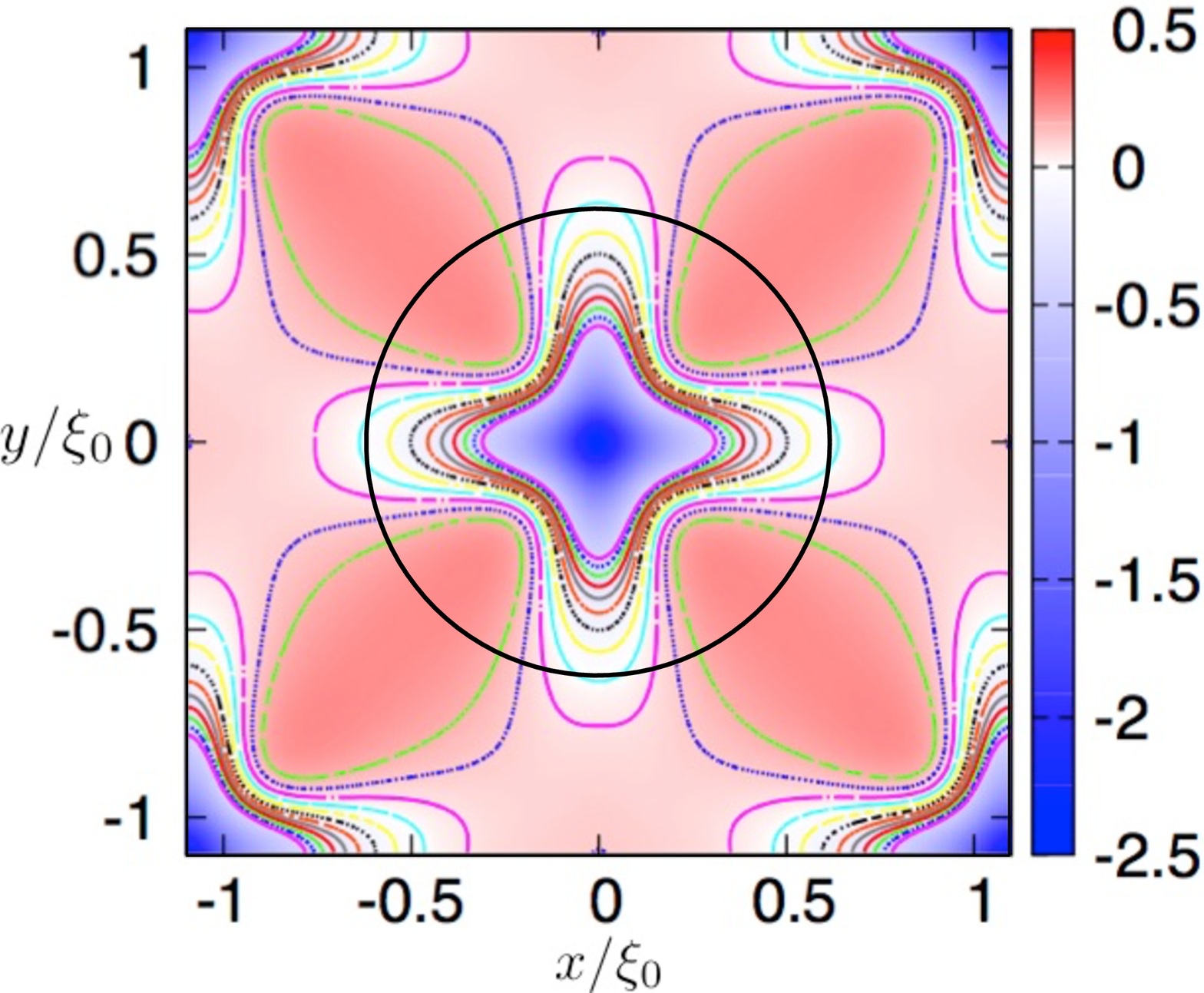}
                \end{center}
\caption{(Color online) Charge density $\rho({\bm r})/\rho_0$ in the vortex lattice calculated for the case of $n=1.95$ and ${{\mu_0\bar{B}/H_{{\rm{c}}2}}}=0.29$ over $-a_2/2\leq x,y \leq a_2/2$, where $a_2/\xi_0=2.2$. }
                \label{fig4}
                \end{figure}
 \begin{figure}[t]
        \begin{center}
                \includegraphics[width=0.9\linewidth]{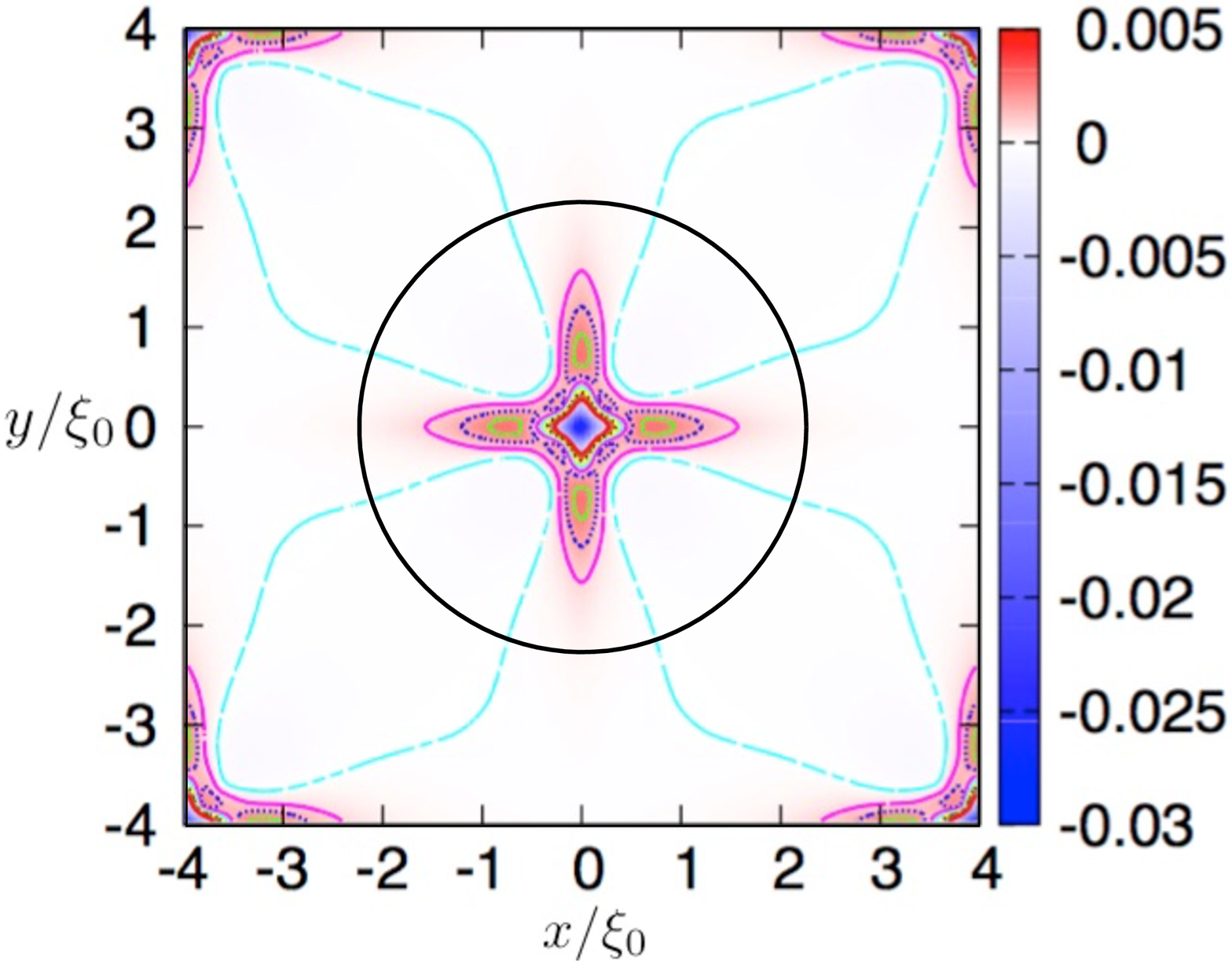}
                \end{center}
                \caption{(Color online) Charge density $\rho({\bm r})/\rho_0$ in the vortex lattice calculated for the case of $n=0.9$ and ${{\mu_0\bar{B}/H_{{\rm{c}}2}}}=0.018$ over $-a_2/2\leq x,y \leq a_2/2$, where $a_2/\xi_0=8.0$.}
                \label{fig5}
                \end{figure}
                \begin{figure}[t]
                \begin{center}
                \includegraphics[width=0.9\linewidth]{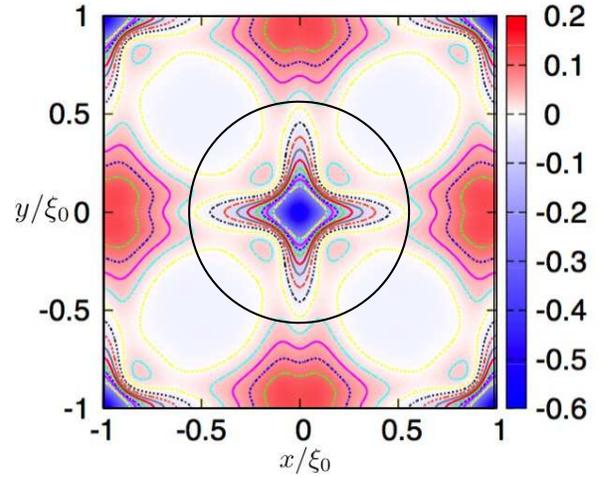}
                \end{center}
                \caption{(Color online) Charge density $\rho({\bm r})/\rho_0$ in the vortex lattice calculated for the case of $n=0.9$ and ${{\mu_0\bar{B}/H_{{\rm{c}}2}}}=0.29$ over $-a_2/2\leq x,y \leq a_2/2$, where $a_2/\xi_0=2.0$.}
                \label{fig6}
                \end{figure}

 Equation (\ref{dete}) with the condition {{in Eq.}} (\ref{BC2}) is an {$H_{{\rm{c}}2}$} {{equation}} {{that}} is applicable to clean superconductors with {{arbitrary}} energy gaps and Fermi surfaces.
Indeed, it reduces to the {{Helfand--Werthamer}} theory\cite{HW} without impurity scattering by setting $\phi$ to $1$ and using the spherical Fermi surface. 
It was numerically found for the present $d_{x^2-y^2}$ model that $N_{\rm{cut}}=12$ yields {{satisfactry}} convergence at all temperatures.
We show the temperature dependence of the reduced critical field $h^{*}\equiv (-\left.dH_{{\rm{c}}2}/dt\right|_{t=1})^{-1}H_{{\rm{c}}2}(t)$ $(t\equiv T/T_{\rm{c}})$ of this model in {{comparison}} with the result {{for}} a two-dimensional $s$-wave superconductor in Fig. \ref{fig2}.
We see that the energy-gap or {{Fermi surface}} anisotropy {{enhances}} the value of $h^{*}$ as  {{the}} temperature is lowered.

Using the solutions thereby obtained by the standard Eilenberger equations, 
we numerically calculated the charge density caused by the Lorentz force in the $d_{x^2-y^2}$ vortex lattice. 
The results presented below were obtained for $T/T_{\rm{c}}=0.3$, $\delta=0.05$, $\lambda_{\rm{TF}}/\xi_0=0.05${{,}} and $\lambda_{\rm L}/\xi_0\equiv\sqrt{\hbar/\mu_0\Delta_0\xi_0e^2 N(0) \langle{v_{\rm{F}}}\rangle_{\rm{F}}}/\xi_0=100.0$, where $\Delta_0$ denotes the energy gap at $(H,T)=(0,0)$. The charge density and total charge in two dimensions are normalized by $\rho_0\equiv\epsilon_0\Delta_0/|e|\xi_0^2$ and $Q_0\equiv\epsilon_0\Delta_0/|e|$, respectively.
We also {{refer to}} the area $r\leq a_{2}/(2\sqrt{\pi})\simeq0.28a_{2}\equiv r_{\rm{in}}$ as the {\it vortex core region} (VCR){{,}} which occupies {{half}} of the unit cell of the vortex lattice, where $r\equiv\sqrt{x^{2}+y^{2}}$.
Circles with a radius of $r=r_{\rm{in}}$ are drawn in all figures below in order to visually indicate {{the}} VCR. 

First, we show the results calculated for $n=1.95${{,}} where the Fermi surface is almost isotropic. 
Figures \ref{fig3} and \ref{fig4} show the spatial distributions of the charge density 
at ${{\mu_0\bar{B}/H_{{\rm{c}}2}}}=0.022$ and $0.29$, respectively, over the range of $-a_2/2\leq x,y \leq a_2/2$.
In {{Fig. \ref{fig3}}}, the fourfold symmetry of the charge density in the vortex core is caused solely by the gap anisotropy since each vortex at
 ${{\mu_0\bar{B}/H_{{\rm{c}}2}}}=0.022$ is almost isolated\cite{Ueki}. 
As the magnetic field is increased further, the vortex lattice symmetry generally starts to affect the charge distribution 
{{owing}} to the overlapping of vortices\cite{WK}.
We observe the effect of overlapping clearly in Fig.\ 4 as compared with {{in}} Fig.\ 3;
however, both the fourfold symmetry and the basic charge profile {{remain}} invariant.
Note that the value of the accumulated charge in {{the}} VCR is found to be always negative as a function of magnetic field.

Figures \ref{fig5} and \ref{fig6} show the charge profiles  at ${{\mu_0\bar{B}/H_{{\rm{c}}2}}}=0.018$ and $0.29$, respectively, 
calculated for the realistic filling of $n=0.9$ for high-$T_{\rm{c}}$ superconductors.
In Fig. \ref{fig5} at ${{\mu_0\bar{B}/H_{{\rm{c}}2}}}=0.018$, the region of the negative charge density extends along the $45^{\circ}$ directions far outside {{the}} VCR {{owing}} to the cooperative effect of the {{energy gap}} and Fermi surface anisotropies.\cite{Ueki}
By increasing the magnetic field, positive charge accumulates at interstitial regions of neighboring vortices {{owing}} to their overlapping,
as shown in Fig. \ref{fig6}, causing {{a}} reduction of the positive charge in the core region.
We point {{out}} that the change {{in}} the charge profile from Fig. \ref{fig5} to Fig. \ref{fig6} is {{much greater than}}
that from Fig.\ 3 to Fig.\ 4 at $n=1.95$.
Its origin can be traced to the competition between the energy-gap and {{Fermi surface}} anisotropies,
{{which}} is far more conspicuous at $n=0.9$.

\begin{figure}[t]
        \begin{center}
                \vspace{4mm}\includegraphics[width=0.94\linewidth]{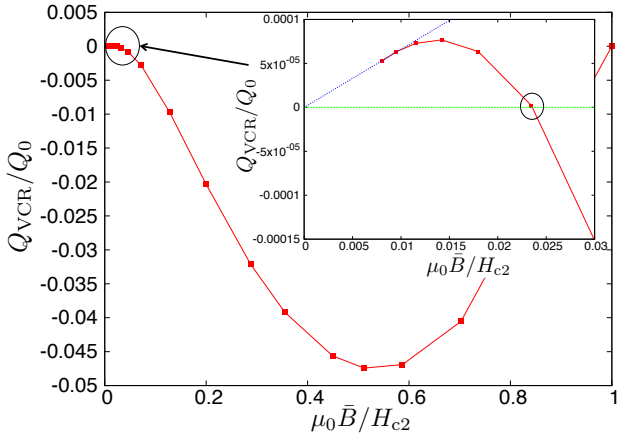}
                \end{center}
\caption{(Color online)
Total charge inside {{the}} VCR, $Q_{\rm{VCR}}/Q_{0}$, as a function of ${{\mu_0\bar{B}/H_{{\rm{c}}2}}}$ calculated for the case of $n=0.9$. The {{sign reversal}} of $Q_{\rm{VCR}}$ can be observed at ${{\mu_0\bar{B}/H_{{\rm{c}}2}}}=0.024$.
}
 \label{fig7}
 \end{figure}
 In figure \ref{fig7}, we show {{the}} magnetic field dependence of the accumulated charge $Q_{\rm{VCR}}\equiv\int_{0}^{2\pi} \int_{0}^{r_{\rm{in}}} r\rho(r,\varphi) drd\varphi$ calculated at $n=0.9$.
 We see that {{$|Q_{\rm{VCR}}|$}} is enhanced strongly by {{the}} magnetic field with a peak near $H_{{\rm{c}}2}/2$.
 This peak structure, which is characteristic of the Lorentz force mechanism for the charging, originates from the competition between the increasing magnetic field and the decreasing pair potential\cite{WK}.
 The absolute value of this peak can be $10^2$--$10^3$ times larger than that for ${{\mu_0\bar{B}/H_{{\rm{c}}2}}}\lesssim0.05$.
 This feature is commonly seen in our previous study on the $s$-wave case with an isotropic Fermi surface. 
 However, for the present anisotropic $d_{x^2-y^2}$ pairing on {{a reasonably}} anisotropic Fermi surface of $n=0.9$,
 we even observe a sign change of $Q_{\rm{VCR}}$ as a function of magnetic field
 as seen in the inset of Fig.\ 7.
      \begin{figure}[t]
        \begin{center}
        \vspace{4mm}\includegraphics[width=0.9\linewidth]{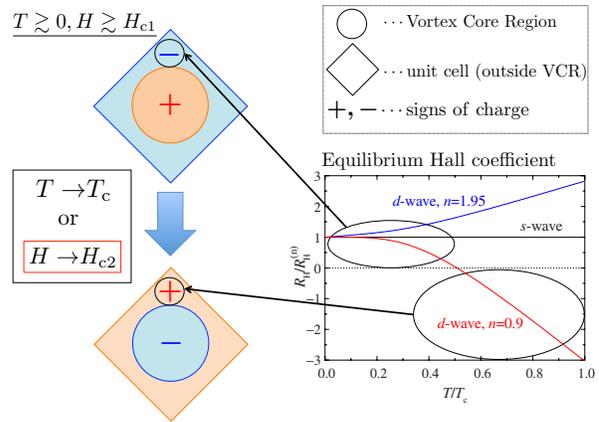}
                \end{center}
\caption{(Color online)
{
Schematic drawing {{of}} the sign reversal of the vortex-core charge for $n=0.9$ using Fig. 3 in Ref. {\citen{Kita09}}.
The normal Hall coefficient $R^{({\rm{n}})}_{\rm{H}}$ is negative for $n=0.9$.
The accumulated charge inside {{the}} VCR, $Q_{\rm{VCR}}$, has the opposite sign to {{that}} outside {{the VCR owing}} to the charge neutrality.
The  accumulated charge outside {{the}} VCR may change its sign because the excitation curvature for $H:H_{{\rm{c}}1} \to H_{{\rm{c}}2}$ changes similarly {{to that}} for $T:0\to T_{\rm{c}}$.\cite{Ueki,Kita09}}
}
 \label{fig8}
 \end{figure}     
{
On the basis of the charge neutrality of the system, we can {{analyze}} the sign of $Q_{\rm{VCR}}$ in terms of the sign of the accumulated charge outside {{the}} VCR.
{{A}} schematic drawing of the sign reversal of the vortex-core charge observed {{for}} $n=0.9$ is shown in Fig. \ref{fig8}. 
Assuming that $|\Delta|$ is constant to the zeroth order, we may describe the electric field outside {{the}} VCR by
\begin{equation}
\bm{E}=\bm{B}\times\underline{R}_{\rm{H}}\bm{j}_{\rm{s}}{{,}}
\end{equation}
where $\underline{R}_{\rm{H}}$ is the equilibrium Hall coefficient tensor in the  Meissner state{{,}}\cite{Kita09,Ueki} 
\begin{equation}
\underline{R}_{\rm{H}}=\frac{1}{2eN(0)} \left\langle\frac{\partial}{\partial \bm{p}_{\rm{F}}} (1-Y)\bm{v}_{\rm{F}}\right\rangle_{\rm{F}}\langle\bm{v}_{\rm{F}} (1-Y)\bm{v}_{\rm{F}}\rangle_{\rm{F}}^{-1},\label{Hall}
\end{equation} 
$\bm{j}_{\rm{s}}$ is {{the}} supercurrent, and $Y$ is the Yosida function\cite{Yosida}.
The Hall coefficient, which determines the sign and magnitude of the {{carriers}} in the Meissner state, has the temperature dependence given in Fig.\ \ref{fig8},
which exhibits {{a}} sign change due to the change {{in}} the excitation curvature under the growing energy gap as $T\rightarrow 0$.
Now, its high-temperature (low-temperature) region may be identified with the high-field (low-field) case in the magnetic-field dependence of the charge accumulation outside {{the}} VCR at $T=0.3T_{\rm c}$.
This identification {{allows us to}} describe the magnetic-field dependence of the charge accumulation {and its} sign change outside {{the}} VCR.
Thus, we may conclude that the sign change seen {{in}} Fig.\ \ref{fig7} as a function of magnetic field is brought about by the change in the excitation curvature under the decreasing pair potential as $\mu_0\bar B\rightarrow H_{{\rm c}2}$.
In this context, we  note that the {{vortex-core}} charge in {{an}} $s$-wave vortex lattice with a circular Fermi surface has the same sign (positive) with increasing magnetic field since the excitation curvature outside {{the VCR}} does not change. 
}
             
In summary, we have numerically studied { the magnetic-field dependence of the vortex-core charge caused by the Lorentz force}, focusing our attention on the competition between the energy-gap and {{Fermi surface}} anisotropies.
We found that the accumulated charge in {{the}} VCR may reverse its sign as a function of magnetic field {{as well as a function of}} temperature.\cite{Ueki}
The sign of the vortex-core charge has been discussed\cite{Khomskii} in connection with the anomalous sign change of the flux-flow Hall conductivity\cite{Hagen93,Nagaoka98,Puica} observed in various type-II superconductors.
Whether the present mechanism for the sign change is relevant to the sign change of the flux-flow Hall conductivity has yet to be clarified,
{{which will require}} a detailed calculation of the Hall coefficient in the resistive flux-flow regime.


\begin{thebibliography}{9}

\bibitem{WK}W. Kohno, H. Ueki, and T. Kita, J. Phys. Soc. Jpn. {\bf{85}}, 083705 (2016).

\bibitem{Kita01}T. Kita, Phys. Rev. B {\bf 64}, 054503 (2001).

\bibitem{Ueki}H. Ueki, W. Kohno, and T. Kita, J. Phys. Soc. Jpn. {\bf{85}}, 064702 (2016).

\bibitem{Ziman}See, for example, J. M. Ziman, {\em Electrons and Phonons} (Oxford University Press, Oxford, 1960), {{p. $486$}}.

\bibitem{Kita09}T. Kita, {Phys. Rev. B} {\bf 79}, 024521 (2009).

\bibitem{Kopnin}N. B. Kopnin, {\it Theory of Nonequilibrium Superconductivity} (Oxford University Press, New York, 2001). 

\bibitem{Parks69}R. D. Parks (ed.), {\em Superconductivity} (Marcel Dekker, New York, 1969).

\bibitem{KitaText}T. Kita, {\it Statistical Mechanics of Superconductivity} (Springer, Tokyo, 2015).

\bibitem{Rainer83}J. W. Serene and D. Rainer, Phys. Rep. {\bf 101}, 221 (1983).

\bibitem{LO86}{A. I. Larkin and Y. N. Ovchinnikov,
in {\em Nonequilibrium Superconductivity}, ed. D. N. Langenberg and A. I. Larkin (Elsevier, Amsterdam, 1986) Vol. 12, p. 493.}

\bibitem{Kontani}H. Kontani, Rep. Prog. Phys. {\bf{71}}, 026501 (2008).

\bibitem{KitaGL}T. Kita, J. Phys. Soc. Jpn. {\bf{67}}, 2067 (1998).

\bibitem{Ichioka1}M. Ichioka, N. Hayashi, N. Enomoto, and K. Machida, {Phys. Rev. B} {\bf 53}, 15316 (1996). 

\bibitem{VFex1}I. Maggio-Aprile, Ch. Renner, A. Erb, E. Walker, and \O. Fischer, Phys. Rev. Lett. {\bf 75}, 2754 (1995).

\bibitem{Ichioka2}M. Ichioka, A. Hasegawa, and K. Machida, Phys. Rev. B {\bf{59}}, 8902 (1999).

\bibitem{VFex2}A. S. Cameron, J. S. White, A. T. Holmes, E. Blackburn, E. M. Forgan, R. Riyat, T. Loew, C. D. Dewhurst, and A. Erb{{.}} Phys. Rev. B {\bf 90}, 054502 (2014).

\bibitem{Klein}U. Klein, {J. Low Temp. Phys.} {\bf 69}, 1 (1987). 

\bibitem{KA04}T. Kita and M. Arai, Phys. Rev. B {\bf 70}, 224522 (2004).

\bibitem{HW}E. Helfand and N. R. Werthamer, Phys. Rev. Lett. {\bf{13}}, 686 (1964).

{{
\bibitem{Yosida}
K. Yosida, Phys. Rev. {\bf{110}}, 769 (1958).}}

\bibitem{Khomskii}D. I. Khomskii and A. Freimuth, Phys. Rev. Lett. {\bf{75}}, 1384 (1995).

\bibitem{Hagen93}For an experimental overview and references, see, for example, 
S. J. Hagen, A. W. Smith, M. Rajeswari,
J. L. Peng, Z. Y. Li, R. L. Greene, S. N. Mao, X. X. Xi,
S. Bhattacharya, Q. Li, and C. J. Lobb, Phys. Rev. B {\bf 47}, 1064 (1993).

\bibitem{Nagaoka98}T. Nagaoka, Y. Matsuda, H. Obara, A. Sawa, T. Terashima, I. Chong, M. Takano, and M. Suzuki, Phys. Rev. Lett. {\bf 80}, 3594 (1998).

\bibitem{Puica}I. Puica, W. Lang, W. G\"ob, and {{R.}} Sobolewski, Phys. Rev. B {\bf{69}}, 104513 (2004).


\end{thebibliography}
\end{document}